\documentclass[preprint,prc,aps,showpacs,showkeys,groupedaddress,floatfix]{revtex4}
\usepackage{epsfig}
\usepackage{dcolumn}
\usepackage{bm}
\usepackage{graphics}
\usepackage{graphicx}
\usepackage{amssymb}
\usepackage{amsmath}
\usepackage{epstopdf}
\usepackage{float}

\begin{document}

\title{Probing the anomalous $tq\gamma$ couplings through single top production at the future lepton-hadron colliders}

\author{E. Alici}
\email[]{edaalici@beun.edu.tr} \affiliation{Department of Physics, Zonguldak Bulent Ecevit University, Turkey}

\author{M. K\"{o}ksal}
\email[]{mkoksal@cumhuriyet.edu.tr} \affiliation{Department of Optical Engineering, Sivas Cumhuriyet University, Turkey}

\begin{abstract}
The measurements of the top quark flavor changing neutral current interactions are one of the most important goals of the top quark physics program in the present and the future collider experiments. These measurements provide direct information on non-standard interactions of the top quark. Within the framework of new physics beyond the Standard Model, these interactions can be defined by an effective Lagrangian. In this study, we have investigated the potential of the future $\mu p$ colliders on the top quark flavor changing neutral current interactions through the subprocesses $\gamma q \rightarrow t \rightarrow W b$ where $q=u,c$. These subprocesses have been produced through the main reaction $\mu p \rightarrow \mu \gamma p \rightarrow \mu W b X  $  at the LHC$-\mu p$, the FCC$-\mu p$ and the SPPC-$\mu p$ . For the main reaction, the total cross sections have been calculated as a function of the anomalous $ tq\gamma $ couplings.
In addition, sensivities on BR($t \rightarrow q \gamma$) at $95\%$ Confidence Level have been calculated. We obtain that the best constraints on BR($t \rightarrow q \gamma$) are at the order of 10$^{-7}$ which is four orders of magnitude better than the LHC's experimental results.

\end{abstract}



\maketitle

\section{Introduction}
	
The Standard Model (SM) that is the fundamental theory of particle physics is a confirmed strong theory experimentally. Especially, with the ultimate discovery of the approximately 125 GeV Higgs boson at the LHC in 2012, the SM has achieved an important success \cite{hh,hh1}. Although the SM has many significant successes, it includes still important questions unanswered. In order to answer these questions, different theoretical models beyond the SM, called new physics (NP), have been suggested in the literature. In addition, the interactions of the top quark take more attention among these NP model interactions, because the top quark is the only fermion that has mass at the scale of electroweak symmetry breaking  \cite{1}.
	
	One of the most attractive NP models in relevant with top quark physics is Flavour Changing Neutral Currents (FCNC) transitions. In the FCNC transitions, the fermions interact with neutral gauge bosons, changing its flavour without changing the electric charge. Such transitional processes in the SM have been forbidden at tree level and strongly suppressed in the one loop level by the Glashow-Iliopoulos-Maiani (GIM) mechanism \cite{2,3}. On the other hand,  there are various models beyond the SM which make possible to FCNC transitions \cite{4,5,6,7,8,9,10,11,12}. One of these is a model-independent effective Lagrangian method which based on the idea that SM is the low energy limit of a more fundamental theory. In this context, it can be used to examine the effects of the top quark FCNC interactions by determining deviations in the SM predictions.
	
The effective Lagrangian that gives rise to FCNC interactions between the top quark, the photon and any of $u$ and $c$ quarks is identified as follows
	
\begin{eqnarray}
{\mathcal{L}}=\sum_{q=u,c} ig_{e}e_{t}\bar{t}\frac{{\sigma_{\mu\nu}p^{\nu}} }{{\Lambda}}{\kappa}_{tq\gamma} q A^{\mu}
\end{eqnarray}
where $\sigma_{\mu\nu}=[\gamma_{\mu},\gamma_{\nu}]/2  $ with $ \gamma_\mu $ which stands for the Dirac matrix, $ \kappa_{tq\gamma}$ is the anomalous coupling constant,  $g_{e}$ is the electromagnetic coupling constant, $e_{t}$ is the electric charge of the top quark. Also, $\Lambda$ is the conventionally taken mass of the top quark in here. Thus, we take $\Lambda=m_{t}$ where $m_{t}$ is the mass of the top quark. Besides, we assume $\kappa_{tu\gamma}=\kappa_{tc\gamma}$ in our calculations.
The anomalous decay width of the top quark to $u$ or $c$ quarks can be written as follows

\begin{eqnarray}
{\Gamma(t{\rightarrow}q\gamma)}=\frac{g_{e}^2e_{t}^2{\kappa}_{tq\gamma}^2m_t^3}{8\pi\Lambda^2}.
\end{eqnarray}
In the above equation, the masses of $u$ and $c$ quarks are omitted.
The major decay channel of the top quark is $t\rightarrow W b$, the decay width for this channel is given as follows \cite{16}

\begin{eqnarray} {\Gamma(t{\rightarrow}bW)=\frac{\alpha|V_{tb}|^{2}}{16s_{w}^{2}}\frac{m_{t}^{3}}{m_{w}^{2}}(1-3m_{W}^{4}/m_{t}^{4}+2m_{W}^{6}/m_{t}^{6})}
\end{eqnarray}
where $|V_{tb}|$ is the CKM matrix element and $ {\alpha}$ is the fine structure constant, $s_{w}$ is defined the sine of the Weinberg angle. Also, $ m_{W}$ is the mass of the W boson. In Eq. 3, we have used as $|V_{tb}|=0.999$, $\alpha=\frac{1}{128.921}$, $m_{t}=172.5$ GeV, $m_{w}=80.4$ GeV and $s_{w}^{2}=0.2342$. Hence, the decay width for $t\rightarrow W b$ have obtained as $\Gamma(t\rightarrow W b)\simeq1.44$ GeV. Therefore, the branching ratio of the anomalous $t \rightarrow q \gamma$ coupling can be generally described as follows

\begin{eqnarray}
{BR(t\rightarrow q\gamma)}=\frac{\Gamma(t\rightarrow q\gamma)}{\Gamma(t\rightarrow W b)}.
\end{eqnarray}

In the literature, there are many experimental and theoretical studies concerned with FCNC interactions of the top quark via effective Lagrangian method \cite{17,18,19,20,21,22,24,25,26,27,28,29,30,31,32,33,34,35,36,37,38,39,40,41,42,43,44,45,46,47,48,49,50,51,52,53,531,532,533,54,55,56,57,58,ch1,ch2}.
Experimental limits in related with the FCNC interactions have been obtained by various particle collider collaborations. Provided by the CDF collaboration, the branching ratio at $95\%  $ Confidence Level (C. L.) for $ t\rightarrow q\gamma $ decay is $BR(t\rightarrow q\gamma)<3.2\%$  \cite{14}. Besides, obtained upper limit at $95\%  $ C.L. on the anomalous $tq\gamma$ couplings by the ZEUS collaboration is $ \kappa_{tq\gamma}<0.12$  \cite{15}. In addition to those mentioned above, the most recent experimental constraints on the top quark FCNC couplings have been obtained  by the CMS collaboration.  These constraints are $BR(t\rightarrow u\gamma)<1.3\times10^{-4}$ and $BR(t\rightarrow c\gamma)<1.7\times10^{-3}$ \cite{59}. Furthermore, by utilizing Eqs. 2-4, the sensitivity limits of the anomalous coupling have been calculated as $\kappa_{tq\gamma}<0.094$.

\section{MUON-PROTON COLLIDERS}

In particle physics, researches aim to get more information about new physics by using developed accelerator technologies which allow different types of collisions such as hadron-hadron, lepton-lepton, lepton-hadron to be examined. The Large Hadron Collider (LHC)  is a circular proton-proton collider which is the most powerful particle accelerator built so far. In parallel to developments in accelerator technology, the LHC will be upgraded and utilized as a lepton-hadron collider in the near future. In this regard, it is first planned that the LHC will be transformed into the Large Hadron electron Collider(LHeC) having an electron ring to be tangentially constructed to the main tunnel of the LHC. After the completion of the LHeC mission, the LHC$-{\mu} {p}$ option will be activated by replacing the electron ring with the muon ring. On the other hand, it is planned to construct two new multi-purpose accelerator complexes with high luminosity at energy frontier in the far future. One of these complexes is the Future Circular Collider (FCC) machine to be built into the post-LHC area at CERN in Europe in 2030s. According to the FCC study reports, it will be mainly used as \textit{pp} collider with 100 TeV center-of-mass energy at 100 km circular tunnel. Besides, it is planned that \textit{ee}, $\mu \mu$,  \textit{ep} and $\mu p$ collisions will be performed, as the additional options \cite{60,61,62,63}. The other complex, Super Proton Proton Collider (SPPC), is planned to be implemented by Chinese scientists at the same time as the FCC \cite{64}. In this context, the SPPC will be activated as a \textit{pp}  collider with 70 TeV centre-of-mass energy at the 100 km main tunnel. Besides, SPPC will also be operated \textit{ee}, $\mu \mu$,  \textit{ep} and $\mu p$ interactions as in other accelerators.

However, the general expression for the luminosity of based muon-proton colliders is given by \cite{saleh}

\begin{eqnarray}
L_{\mu p}=\frac{N_{\mu}N_{p}}{4\pi max [\sigma_{x_{p}},\sigma_{x_{\mu}}]max [\sigma_{y_{p}},\sigma_{y_{\mu}}]} min[f_{c_{p}}f_{c_{\mu}}]
\end{eqnarray}
where $N_{\mu}$ and $N_{p}$ are numbers of muons and protons per bunch, respectively; $\sigma_{x_{p}}$ ($\sigma_{x_{\mu}}$) and $\sigma_{y_{p}}$ ($\sigma_{y_{\mu}}$) are the horizontal and vertical proton (muon) beam sizes at interaction point; $f_{c_{p}}$ and $f_{c_{\mu}}$ are bunch frequencies. $f_{c}$ is described by $f_{c}=N_{b}f_{rev}$, where $N_{b}$ is number of bunches, $f_{rev}$ means revolution frequency.

Moreover, muon-proton colliders  can be used not only as $\mu p  $ colliders but also as $\mu \gamma $, $\gamma \gamma $ and $\gamma p $ colliders. These photons can be real photons that fit the compton backscattering mechanism or that can be quasi-real photons conform to the Equivalent Photon Approximation (EPA) which is a more realistic approach than the compton backscattering mechanism.
In the EPA, incoming muon or proton are scattered with the very small angle from beam pipe by losing a small part of their transverse momentum and by emitting a photon at forward direction \cite{65,66,67}. Firstly, these emitted photons either interact with each other and achieve $\gamma \gamma $ collisions. Secondly,
a quasi-real photon emitted from the incoming muon beam can interact with proton beam shortly and thus $\gamma p $ processes occur.
In this respect, the main parameters for $\mu p$ colliders, used in examining the process of $\gamma p $ interaction in our study, are presented in Table I \cite{68,69,70,71}.
	
In this study, we have investigated the anomalous $t q \gamma$ couplings for the process $\mu p \rightarrow \mu \gamma p \rightarrow \mu W b X$ with the subprocesses $\gamma q \rightarrow t \rightarrow W b$  where $q=u,c$. Also, we have obtained 95\% C.L. limits on $BR(t \rightarrow q\gamma)$ as a function of integrated luminosities under different configurations of center-of-mass energy for the LHC-$\mu p$, the FCC-$\mu p$ and the SPPC-$\mu p$.

\section{CROSS SECTIONS AND SENSITIVITY ANALYSIS}

In this section, we have examined the potential of the future $\mu p$ colliders on the anomalous FCNC transitions through ${tq\gamma}$ interactions.
Schematic diagram for the process $\mu p \rightarrow \mu \gamma p{\rightarrow}\mu W b X$ is shown Fig. 1. The tree level Feynman diagrams for the subprocesses $\gamma q\rightarrow W b$ in our calculations are displayed in Fig. 2. These subprocesses include four Feynman diagrams. Three of these diagrams give only SM contribution and a Feynman diagram consists of the anomalous ${tq\gamma}$ contribution arising from new physics.
For the main process $\mu p \rightarrow \mu \gamma p{\rightarrow}\mu W b X$, the total cross sections as a function of the anomalous $ \kappa_{tq\gamma}$ couplings for various center-of-mass energy values of the LHC-$\mu p$, the FCC-$\mu p$ and the SPPC-$\mu p$ have been investigated. We have performed numerical calculations on cross sections by using the CalcHEP simulation program, in which new interaction vertices has been added \cite{72}.  In here, the distribution function of photon emitted by muon in the EPA is given by

\begin{eqnarray}
{f_{\gamma}(x)=\frac{\alpha}{\pi E_{\mu}}\{[\frac{1-x+x^{2}}{x}]log(\frac{Q_{max} ^ {2}}{Q_{min} ^ {2}})-\frac{m_{\mu}^{2}x}{Q_{min} ^ {2}}(1-\frac{Q_{max} ^ {2}}{Q_{min} ^ {2}})-\frac{1}{x}[1-\frac{x}{2}]^{2}log(\frac{x^{2}E_{\mu}^{2}+Q_{max} ^ {2}}{x^{2}E_{\mu}^{2}+Q_{min} ^ {2}})\}}
\end{eqnarray}
where $x=\frac{E_{\gamma}}{E_{\mu}}$ and $Q_{max}^{2}$ is the maximum virtuality of the photon. During the calculation, the photon virtuality is taken as $Q_{max} ^ {2} = 2$ GeV$^{2}$. The minimum value of $Q_{min}^{2}$ is given as follows

\begin{eqnarray}
{Q_{min} ^ {2}=\frac{m_{\mu}^{2}x}{1-x}}
\end{eqnarray}
where $m_{\mu}$ is the mass of the muon. In addition, for the final state \textit{b} quark, we apply cuts as $ p_{t}> 20$ GeV for the transverse momentum and as $|\eta |<2.5$ for pseudorapidity. For parton distribution functions, the CTEQ6L1 is adopted \cite{73}. The total cross sections of the process $\mu p \rightarrow \mu \gamma p{\rightarrow}\mu W b X$ with respect to the anomalous $ \kappa_{tq\gamma}$ couplings for different center-of-mass energies of the LHC-$\mu p$, the FCC-$\mu p$ and the SPPC-$\mu p$ have been presented in Figs. 3-5. As seen in these figures, the total cross section increases with increasing values of center-of-mass energies. This situation is an expected result due to the new physics contributions having more energy dependence than the SM. In addition, the increasing trend for the total cross section has also been observed for the increasing $ \kappa_{tq\gamma}$ values.

On the other hand, it is necessary to perform statistical analysis to determine whether the effects of the anomalous $ \kappa_{tq\gamma}$ couplings on the cross sections are experimentally measurable. Therefore, we have examined the precision of the process $\mu p \rightarrow \mu \gamma p{\rightarrow}\mu W b X$ to the branching ratios of $t\rightarrow q\gamma$ with the $\chi^{2}$ statistical analysis without a systematic error. The  $\chi^{2}$ function is given by

\begin{eqnarray}
{\chi^{2}=(\frac{\sigma_{sm}-\sigma_{NEW}}{\sigma_{sm}\delta})^{2}}
\end{eqnarray}
where $\sigma_{NEW}$ is the total cross section of the process containing both the SM and the NP contributions and $ \delta=\frac{1}{\sqrt{N}} $ is the statistical error; $N=L_{int}\times BR \times \sigma_{SM}$ represents the number of SM events. In here, BR is the branching ratio of the W boson. We think of both leptonic and hadronic decay of  the W boson in the final state of the process. For this reason, we take $BR(W\rightarrow \ell \nu_{\ell}; \ell=e,\mu) = 0.213$ for leptonic decay and $BR(W\rightarrow q q') = 0.676$ for hadronic decay. In Figs. 6-8, we present sensitivity limits on $BR(t\rightarrow q \gamma)$ for different center-of-mass energies as a function of integrated luminosities of the LHC-$\mu p$, the FCC-$\mu p$ and the SPPC-$\mu p$. In here, we plot Figs. 6-8 by assuming that the W boson decays via leptonic and hadronic channels. We understand from these figures that the hadronic decay channel is a little more sensitive to new physics than leptonic decay channel. Even though, these branching ratios for hadronic and leptonic decay channel are at the same order magnitude with each other.

An important part of our analysis is the inclusion of theoretical uncertainties as there may be several experimental and systematic uncertainty sources in top quark identification. In our study, we consider the total systematic uncertainties of $0,3,5\%$. For leptonic and hadronic decay channels, limits at $95\%$ on the anomalous $\kappa_{tq\gamma}$ coupling through the process $\mu p \rightarrow \mu \gamma p \rightarrow \mu W b X$ at the LHC$-\mu$3000, the FCC$-\mu$1500 and the SPPC2-$\mu$1500,  as well as for various integrated luminosity systematic errors of $0,3,5\%$ are given in Tables II-IV. Tables VII-X represent that the bounds with increasing $ \delta_{sys}$ values at the FCC-he are almost unchanged with respect to the luminosity values and for the center-of-mass energy values. The reason of this situation is $ \delta_{stat}$ which is much smaller than $ \delta_{sys}$.

For the leptonic and hadronic decay channels, the sensitive limits on $BR(t\rightarrow q \gamma)$ at $95\%$ Confidence Level as a function of the integrated luminosity through the process $\mu p \rightarrow \mu \gamma p \rightarrow \mu W b X$ at the LHC$-\mu p$, the FCC$-\mu p$ and the SPPC-$\mu p$ are given in Figs. 6-8. As seen in Fig. 6, the limits on $BR(t\rightarrow q \gamma)$ at the LHC$-{\mu}1500$ with luminosity value of 1 fb$^{-1}$ can set more stringent sensitivities by three orders of magnitude with respect to current experimental limits. The values of branching ratio obtained for the leptonic channel are weaker by a factor of 0.75 than those corresponding to the hadronic channel as shown in Figs. 6 and 7. $\gamma p$ collisions at the 3.5 TeV FCC with an integrated luminosity of 0.02 fb$^{-1}$ represent that the limits on the branching ratio are calculated as $BR(t\rightarrow q \gamma)=1.75\times$ 10$^{-4}$. It improves approximately the sensitivity of branching ratio by up to roughly 10 times compared to the LHC. For the leptonic and hadronic channels, the FCC$-{\mu}1500$ with $\sqrt{s} = 17.3$ TeV and L = 5 fb$^{-1}$ probes the $BR(t\rightarrow q \gamma)$ with a far better than the experiments limits. However, in Fig. 8, we have acquired the most stringent limits on $BR(t\rightarrow q \gamma)$ for the luminosity value of 43 $fb^{-1}$ at the SPPC2$-{\mu}1500$. These limits are four orders of magnitude better than the sensitivity of the LHC. In addition, the best sensitivities derived on $BR(t\rightarrow q \gamma)$ from the process $\mu p \rightarrow \mu \gamma p \rightarrow \mu W b X$ at the LHC$-\mu p$, the FCC$-\mu p$ and the SPPC-$\mu p$ change from the order of 10$^{-4}$ to the order of 10$^{-7}$.

\section{CONCLUSIONS}

The FCNC interactions of the top quark are one of the ways of new physics research beyond the SM. However, the anomalous $tq\gamma$ couplings via $\gamma p$ collisions might also be uniquely revealed in single top quark production. $\mu p$ colliders that are thought to be constructed in the future years can be designed as a high energy $\gamma p$ and $\gamma \gamma$ collider. In this motivation, we have investigated the process $\mu p \rightarrow \mu \gamma p{\rightarrow}\mu W b X$  with the anomalous $tq\gamma$ couplings in a model independent effective Lagrangian approach. Within this framework, we have obtained constraints on $BR(t\rightarrow q\gamma)$ for different center-of-mass energies and integrated luminosities of the LHC-$\mu p$, the FCC-$\mu p$ and the SPPC-$\mu p$. Also, the leptonic and hadronic decay channels of the W boson in the final state of the examined process are considered to detect sensitivities of the anomalous couplings at $\mu p$ colliders. Our results show that the sensitivity to $BR(t\rightarrow q\gamma)$ is three orders of magnitude better than the LHC experimental limits. Consequently, we have inferred that the future $\mu p$ colliders will be playing a key role in the examination of the anomalous interactions of top quark via FCNC transitions.

\begin{table}
  \centering
   \caption{The main parameters of muon-proton colliders}\label{b}
  \begin{tabular}{p{3cm}p{3cm}p{3cm}p{3cm}}
     \hline \hline
     Collider& $E_\mu$(TeV) & $\sqrt{s}$(TeV) & $ \textit{L}(fb^{-1})$ \\
     \hline \hline
      LHC-$\mu$750 & 0.75 & 4.58 & 14 \\
     LHC-$\mu$1500 & 1.5 & 6.48 & 23 \\
     LHC-$\mu$3000 & 3.0 & 9.16 & 9 \\
          \hline \hline
     FCC-$\mu$63 & 0.063 & 3.50 & 0.02 \\
    FCC-$\mu$750 & 0.75 & 12.2 & 5 \\
     FCC-$\mu$1500 & 1.5 & 17.3 & 5 \\
          \hline \hline
     SPPC1-$\mu$750 & 0.75 & 10.33 & 5.5 \\
     SPPC2-$\mu$750 & 0.75 & 14.28 & 12.5 \\
      SPPC1-$\mu$1500 & 1.5 & 14.61 & 4.9 \\
      SPPC2-$\mu$1500 & 1.5 & 20.2 & 42.8 \\
     \hline \hline
   \end{tabular}
\end{table}

\begin{table}
\caption{For systematic errors of $0,3,5\%$, limits on the anomalous $\kappa_{tq\gamma}$ coupling at LHC-$\mu$3000 with various integrated luminosities.}
\begin{center}
\begin{tabular}{|cc|cc|cc|}
\hline
\multicolumn{6}{|c|}{LHC-$\mu$3000} \\
\hline
\multicolumn{2}{|c|}{} & \multicolumn{2}{c|}{Hadronic Channel} & \multicolumn{2}{c|}{Leptonic Channel} \\
\hline
\cline{1-6}
${\cal L} \, (fb^{-1})$  & \hspace{0.5cm} $ \delta_{sys}$ \hspace{0.5cm}  &
\hspace{1.5cm} $\kappa_{tq\gamma}$ \hspace{1.5cm} &&
\hspace{1.5cm} $\kappa_{tq\gamma}$ \hspace{1.5cm} & \\
\hline
1  &  $0\%$   & 0.00474   && 0.00633&   \\
1  &  $3\%$   & 0.00474   && 0.00633&   \\
1  &  $5\%$   & 0.00475   && 0.00633&   \\
\hline
3  &  $0\%$   & 0.00360   &&0.00481&   \\
3  &  $3\%$   & 0.00361   &&0.00481&   \\
3  &  $5\%$   & 0.00362   &&0.00482&   \\
\hline
5  &  $0\%$   & 0.00317   &&0.00423&  \\
5  &  $3\%$   & 0.00318   &&0.00424&  \\
5  &  $5\%$   & 0.00320   &&0.00424&  \\
\hline
7  &  $0\%$   & 0.00291   &&0.00389&   \\
7  &  $3\%$   & 0.00293   &&0.00389&   \\
7  &  $5\%$   & 0.00295   &&0.00391&   \\
\hline
9  &  $0\%$   & 0.00273   &&0.00365&  \\
9  &  $3\%$   & 0.00275   &&0.00366&  \\
9  &  $5\%$   & 0.00278   &&0.00367&  \\
\hline
\end{tabular}
\end{center}
\end{table}

\begin{table}
\caption{Same as in Table II, but for FCC-$\mu$1500.}
\begin{center}
\begin{tabular}{|cc|cc|cc|}
\hline
\multicolumn{6}{|c|}{FCC-$\mu$1500} \\
\hline
\multicolumn{2}{|c|}{} & \multicolumn{2}{c|}{Hadronic Channel} & \multicolumn{2}{c|}{Leptonic Channel} \\
\hline
\cline{1-6}
${\cal L} \, (fb^{-1})$  & \hspace{0.5cm} $ \delta_{sys}$ \hspace{0.5cm}  &
\hspace{1.5cm} $\kappa_{tq\gamma}$ \hspace{1.5cm} &&
\hspace{1.5cm} $\kappa_{tq\gamma}$ \hspace{1.5cm} & \\
\hline
1  &  $0\%$   & 0.00423   &&0.00569&   \\
1  &  $3\%$   & 0.00427   &&0.00569&   \\
1  &  $5\%$   & 0.00428   &&0.00570&   \\
\hline
2  &  $0\%$   & 0.00358   &&0.00478&   \\
2  &  $3\%$   & 0.00359   &&0.00479&   \\
2  &  $5\%$   & 0.00361   &&0.00480&   \\
\hline
3  &  $0\%$   & 0.00324   &&0.00432&  \\
3  &  $3\%$   & 0.00325   &&0.00433&  \\
3  &  $5\%$   & 0.00328   &&0.00434&  \\
\hline
4  &  $0\%$   & 0.00301   &&0.00402&   \\
4  &  $3\%$   & 0.00303   &&0.00403&   \\
4  &  $5\%$   & 0.00306   &&0.00404&   \\
\hline
5  &  $0\%$   & 0.00285   &&0.00380&  \\
5  &  $3\%$   & 0.00287   &&0.00381&  \\
5  &  $5\%$   & 0.00290   &&0.00383&  \\
\hline
\end{tabular}
\end{center}
\end{table}

\begin{table}
\caption{Same as in Table II, but for SPPC2-$\mu$1500.}
\begin{center}
\begin{tabular}{|cc|cc|cc|}
\hline
\multicolumn{6}{|c|}{SPPC2-$\mu$1500} \\
\hline
\multicolumn{2}{|c|}{} & \multicolumn{2}{c|}{Hadronic Channel} & \multicolumn{2}{c|}{Leptonic Channel} \\
\hline
\cline{1-6}
${\cal L} \, (fb^{-1})$  & \hspace{0.5cm} $ \delta_{sys}$ \hspace{0.5cm}  &
\hspace{1.5cm} $\kappa_{tq\gamma}$ \hspace{1.5cm} &&
\hspace{1.5cm} $\kappa_{tq\gamma}$ \hspace{1.5cm} &  \\
\hline
5  &  $0\%$   & 0.00276   &&0.00369&   \\
5  &  $3\%$   & 0.00279   &&0.00370&   \\
5  &  $5\%$   & 0.00283   &&0.00372&   \\
\hline
15  &  $0\%$   & 0.00210   &&0.00281&   \\
15  &  $3\%$   & 0.00215   &&0.00283&   \\
15  &  $5\%$   & 0.00224   &&0.00287&   \\
\hline
25  &  $0\%$   & 0.00185   &&0.00247&  \\
25  &  $3\%$   & 0.00192   &&0.00250&  \\
25  &  $5\%$   &  0.00204   &&0.00256&  \\
\hline
35  &  $0\%$   & 0.00170  &&0.00227&   \\
35  &  $3\%$   & 0.00179  &&0.00231&   \\
35  &  $5\%$   & 0.00193  &&0.00238&   \\
\hline
42  &  $0\%$   & 0.00159   &&0.00213&  \\
42  &  $3\%$   & 0.00170   &&0.00218&  \\
42  &  $5\%$   & 0.00186   &&0.00226&  \\
\hline
\end{tabular}
\end{center}
\end{table}

	\begin{figure}
  \centering
  \includegraphics[width=8.cm]{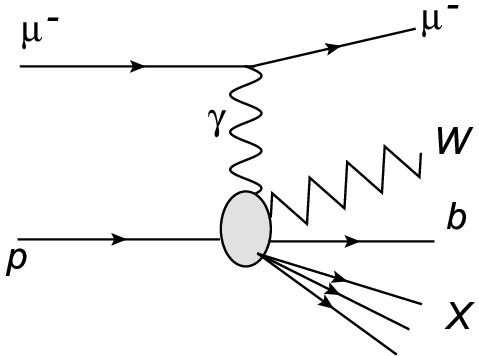}
  \caption{Schematic diagram for the process $\mu p \rightarrow \mu \gamma p{\rightarrow}\mu W b X$}\label{FIG1.}
\end{figure}
\begin{figure}
  \centering
  \includegraphics[width=16.cm]{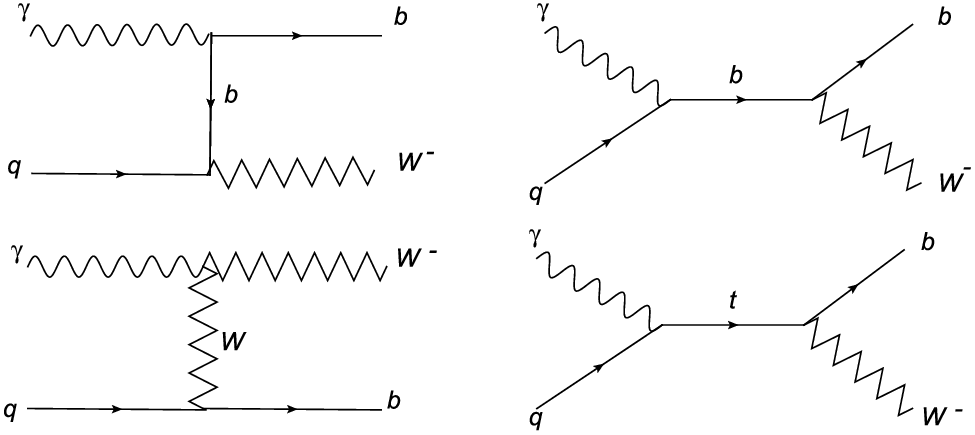}
  \caption{Tree level Feynman diagrams for the subprocesses $\gamma q\rightarrow W b (q=u,c) $  in the presence of the anomalous  $tq\gamma$ couplings. }\label{FIG2.}
\end{figure}
\begin{figure}
  \centering
  \includegraphics[width=10.cm]{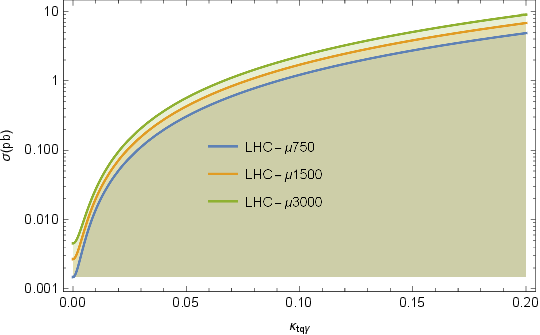}
  \caption{The total cross sections of the process $\mu p \rightarrow \mu \gamma p{\rightarrow}\mu W b X$ as a function of the anomalous ${\kappa}_{tq\gamma}$ coupling for three different center-of mass energies of the LHC-$\mu p$.}\label{FIG9.}
\end{figure}
\begin{figure}
  \centering
  \includegraphics[width=10.cm]{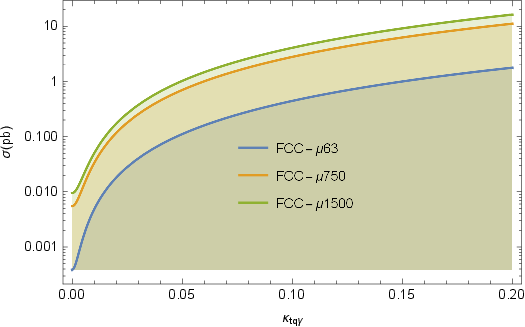}
  \caption{The total cross sections of the process $\mu p \rightarrow \mu \gamma p{\rightarrow}\mu W b X$ as a function of the anomalous ${\kappa}_{tq\gamma}$ coupling for three different center-of mass energies of the FCC-$\mu p$.}\label{FIG9.}
\end{figure}

\begin{figure}
  \centering
  \includegraphics[width=10.cm]{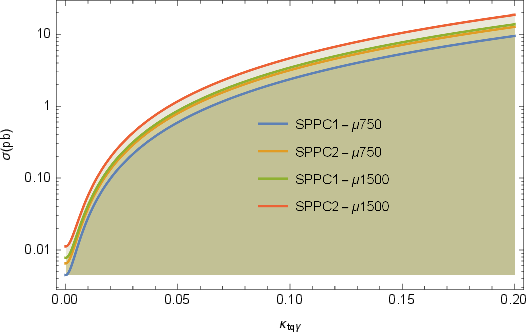}
  \caption{The total cross sections of the process $\mu p \rightarrow \mu \gamma p{\rightarrow}\mu W b X$ as a function of the anomalous ${\kappa}_{tq\gamma}$ coupling for four different center-of mass energies of the SPCC-$\mu p$.}\label{FIG9.}
\end{figure}

\begin{figure}
  \centering
  \includegraphics[width=17.cm,height=7.cm]{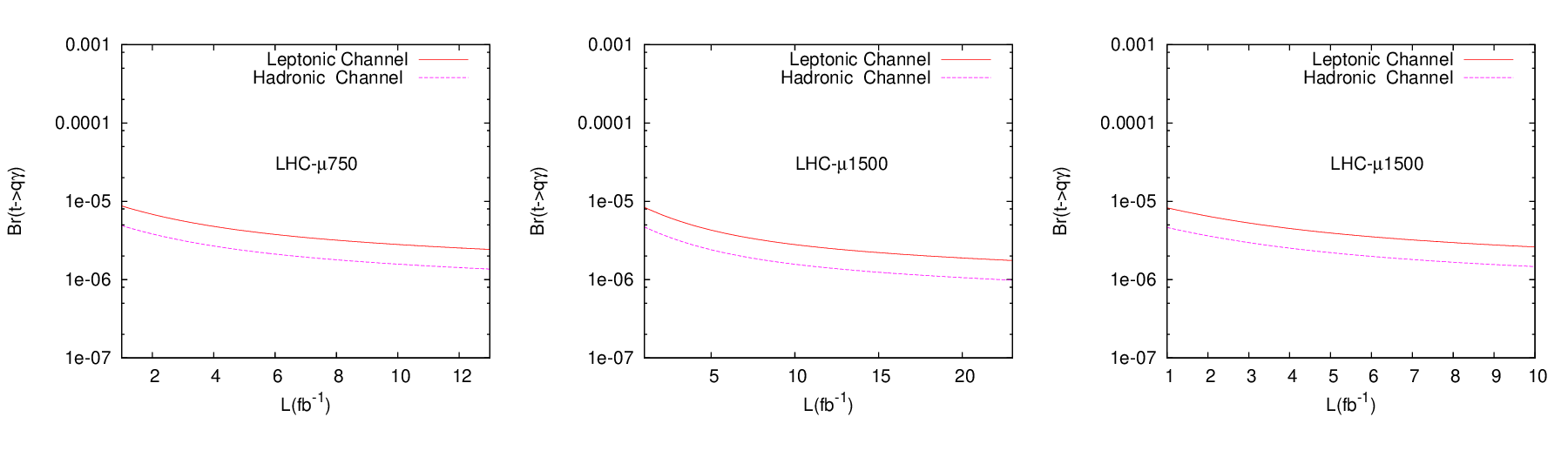}
\caption{For leptonic and hadronic decay channels of the W boson, 95\% C.L. sensitivity limits on $BR(t \rightarrow q\gamma)$
for various integrated luminosities LHC$-{\mu} {p}$.}
\end{figure}

\begin{figure}
\includegraphics[width=17cm,height=7cm]{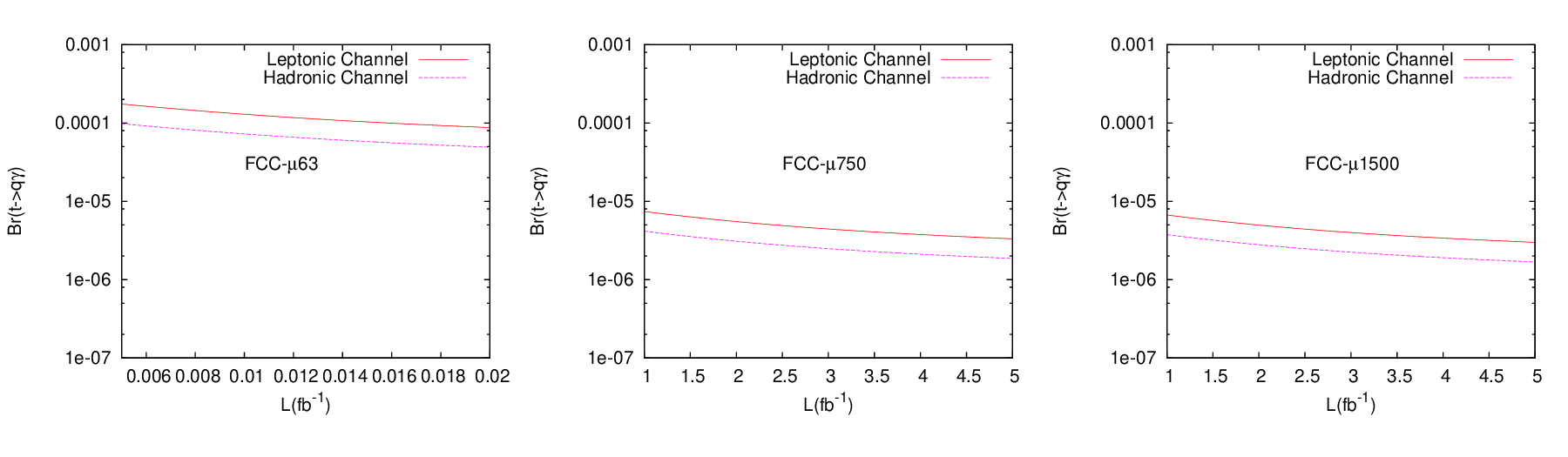}
\caption{For leptonic and hadronic decay channels of the W boson, 95\% C.L. sensitivity limits on $BR(t \rightarrow q\gamma)$
for various integrated luminosities FCC$-{\mu} {p}$.}
\end{figure}

\begin{figure}
\includegraphics[width=17cm,height=7cm]{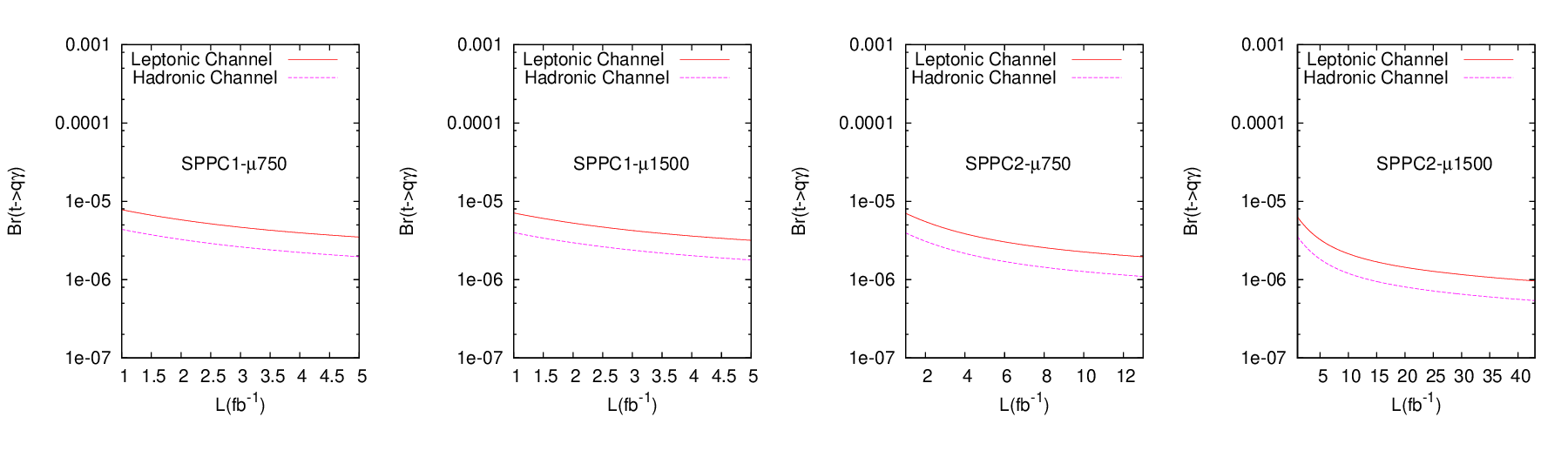}
\caption{For leptonic and hadronic decay channels of the W boson, 95\% C.L. sensitivity limits on  $BR(t \rightarrow q\gamma)$
for various integrated luminosities SPPC$-{\mu} {p}$.}
\end{figure}

\end{document}